# Rapid Oscillations in (TMTSF)$_2$PF$_6$

A.V. Kornilov[*], V.M. Pudalov[*], A.-K. Klehe[#], A. Ardavan[#],
J.S. Qualls[†], J. Singleton[#‡]

[*]*Lebedev Physical Institute, 53 Leninskii prospekt, Moscow, 119991 Russia*
[#]*Clarendon Laboratory, Oxford University, OX1 3PU, UK*
[†]*The University of Texas Pan American, Edinburg, Texas, TX 785390, USA*
[‡]*National High Magnetic Field Lab., LANL, Los Alamos NM 87545, USA*

*In order to clarify the origin of the "Rapid Oscillation" (RO) in (TMTSF)$_2$PF$_6$, we studied the magnetoresistance anisotropy in the Field-Induced Spin Density Wave (FISDW) phase. We have found that in the FISDW insulating state, the Fermi surface is not totally gapped; the remaining 2D metallic pockets are quantized in magnetic field and give rise to the RO. Decreasing temperature does not change the size and orientation of the closed pockets, rather, it causes depopulation of the delocalized states in favor of the localized ones, resulting in the disappearance of the RO.*

## 1. INTRODUCTION

Bechgaard salt (TMTSF)$_2$PF$_6$ demonstrates highly anisotropic conduction, $\sigma_{xx} : \sigma_{yy} : \sigma_{zz} \sim 10^5 : 10^3 : 1$, along ***a***, ***b'***, and ***c**** crystal directions, respectively (for reviews, see Refs. [1-4]). The Fermi surface (FS) in this compound consists of two sheets directed perpendicular to ***a***, which are slightly corrugated due to a small transfer integrals in ***b*** and ***c*** directions. Owing to its quasi-one-dimensionality, the electron system undergoes a phase transition to an insulating spin-density wave (SDW) state at *T* < 12 K.

One of the most interesting features of the SDW phase is the oscillatory magnetoresistance periodic in the inverse field 1/*B*, the so called "Rapid Oscillations" (RO) [5]. Their existence in the insulating phase and the disappearance at low temperatures represent a puzzle. Rapid oscillations have been extensively studied for various Bechgaard salts both experimentally [5-9] and theoretically [10-15]; however, so far no satisfactory and unified understanding of this phenomenon is achieved.

Pressure >0.6GPa suppresses the onset of the SDW and stabilizes a metallic state [3,4]. Strong magnetic field applied along ***c**** restores insulating spin-ordered state either directly from metallic state (at high *T*), or via a cascade of field-induced SDW states [16-18] (at low *T*). RO were earlier reported to exist in the last Field-Induced Spin Density Wave phase (FISDW N=0) [16,17], however, no experimental study followed. To get new insight on the old-standing puzzle of the RO, we studied anisotropy of the oscillations (relative to the magnetic field direction), their *T*-dependence and the domains of their existence in the metallic and FISDW phases. To explore the magnetotransport anisotropy under pressure we rotated in situ the spherical pressure cell with the sample inside [19].

## 2. EXPERIMENTAL RESULTS

Figure 1 shows magnetic field dependence of $R_{xx}$ measured along *a*, with field *B* directed along *c**, i.e. perpendicular to the conducting *a-b* plane.

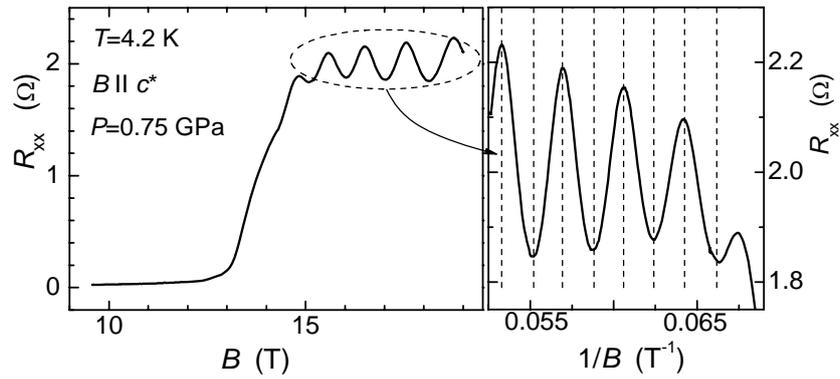

Fig. 1. Magnetoresistance $R_{xx}$ versus magnetic field *B* at *T*=4.2K and *P*=0.75GPa. Right panel blows up the oscillatory part of $R_{xx}(B)$ curve versus inverse field; vertical lines are equidistant in 1/*B*.

As field increases, the resistance initially grows weakly and monotonically, then at $B \approx 13$ T, it sharply raises by a factor of 50, indicating the onset of the FISDW insulating (*N*=0) state, in accord with the known phase diagram [3]. Right after the sharp raise, there are clearly seen oscillations (RO) in $R_{xx}$, which are periodic in 1/*B*. The magnitude of oscillations, starting from the first one (at 14T), sets in a step-like fashion and reaches ~20% at 18T. Would the oscillations persist in the metallic (low field) state, their magnitude extrapolated to the low field have been amounted to about 3% at 12T; however, no oscillations are seen in $R_{xx}$ at *B*<13.5 T. Our data taken at various temperatures, magnetic fields and pressures confirm that RO in

$(TMTSF)_2PF_6$ are intrinsic only to the FISDW state, and hence, are caused by spin ordering.

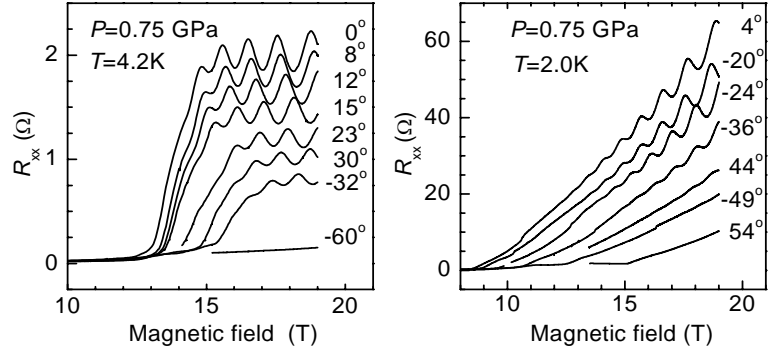

Fig. 2. $R_{xx}(B)$ at $P$=0.75GPa for various magnetic field orientations: for $T$=4.2K (left), and for $T$=2.0K (right). Angle $\theta$ between $c^*$-axis and $B$ is shown next to each curve. $\theta = 0$ corresponds to $B \parallel c^*$. The curves on the left panel are scaled individually (the scaling factors are ×1, ×0.9, ×0.8, ×0.7, ×0.6, ×0.5, ×0.4, ×1, from top to bottom).

When the field is tilted from the normal to the conducting plane, the onset of the FISDW ($N$=0) state shifts to higher fields and the frequency of oscillations increases; examples of $R_{xx}(B)$-curves at $T$=4.2 K and 2 K for tilting the field in the $c^*$-$b'$ plane are shown in Fig. 2. For $P$ = 0.75 GPa, the oscillation frequency equals to 275 T for the field perpendicular to the conducting plane and changes as $275/\cos(\theta)$ Tesla with tilting the field (see Fig. 3). Similar periodic in $1/B$ oscillations with $275/\cos(\theta)$-dependences have been obtained for tilting the field from $c^*$-axes in other planes.

Two important conclusions follow from these observations. (i) The existence of the magnetooscillations indicates that not all electron states are localized [20-22] and large metallic pockets still remain on the FS in the would-to-be insulating phase. The periodicity of oscillations in $1/B$ evidences that RO originate from the quantization of the pockets. (ii) The $1/\cos(\theta)$-dependence proves that these metallic pockets are flat and lie in the $a$-$b$ plane.

The development of RO with temperature is illustrated in Fig. 3 (right panel). As $T$ decreases, the amplitude of oscillations raises, reaches a maximum at ≈3 K and sharply falls down. Nevertheless, neither frequency nor phase of oscillations changes with temperature.

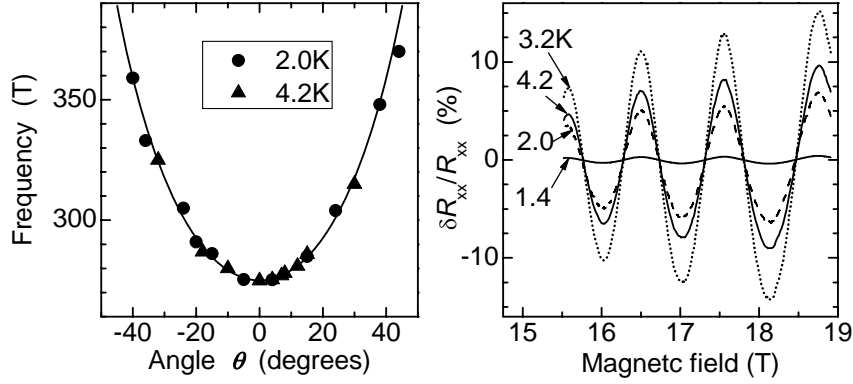

Fig. 3. Left panel: angular dependence of the oscillation frequency for $T$=4.2K (triangles) and 2K (dots). Continuous curve depicts the 275/cos($\theta$)-dependence. Right panel: oscillatory component of the $R_{xx}(B)$ curve at four temperatures; $P$=0.75GPa, $\boldsymbol{B}\|\boldsymbol{c}^*$.

## 3. DISCUSSION

The experimental results described above fit best of all the model suggested by Lebed [11]. In this model, two spin waves exist simultaneously in the spin-ordered state (Fig. 4). Whereas the main wave $\boldsymbol{Q_0}$ localizes electrons on the FS, the auxiliary wave $\boldsymbol{Q_1}$ gives birth to the metallic pockets. The auxiliary wave appears due to Umklapp processes: $\boldsymbol{Q_1} = \boldsymbol{Q_0} - 2\pi/a = \boldsymbol{Q_0} - 4k_F$ in a system with two phase-shifted warped Fermi contours [23,24]; as a result, the wave has a small amplitude. The model requires a commensurability of only $x$-component of the nesting vector with the lattice.

In the model, the RO arise only in the spin-ordered phase as a result of the coexistence of the two spin waves; this is in accord with our observations, Figs. 1 and 2. The model explains the existence of the closed orbits ($a$-$b$-$c$-$d$-$a$), shown in Fig. 4. The closed pockets in the model are two-dimensional and lie in the $a$-$b$ plane; this agrees with the measured 1/cos – angular dependence, Fig.3. The size of the closed pockets in theory depends only on the warping of the FS (i.e. on the $t_b$ transfer integral) and is independent of temperature and magnetic field; this is consistent with our experiment, Figs. 2 and 3. And, at last, the frequency of oscillations calculated on the basis of the model, $4t_b/(\pi e b v_F)$, equals to 286 T, where $e$ is the elementary charge, $b$=6.7Å, $t_b$=200 K, and $v_F$=1.11×10$^5$m/s (as follows from the cyclotron resonance measurements [25]). The experimentally measured frequency 275 T (see Fig. 3) is very close to the calculated value.

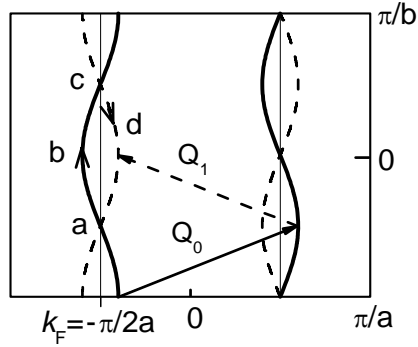

Fig. 4. Schematic view of the first Brillouin zone with the corrugated open FS (two bold lines). Thick arrow shows the main nesting vector $Q_0$, dashed arrow - the auxiliary nesting vector, involving the Umklupp processes $Q_1 = Q_0 - 2\pi/a = Q_0 - 4k_F$ [11]. Arrows show direction of motion in a magnetic field perpendicular to the *a-b* plane.

Thus, the theory explains qualitatively and, in part, quantitatively, *almost all* our experimental results. In order to fit *all* our experimental data, the theory should be accomplished with a mechanism that provides the effectiveness of the Umklapp processes at high temperatures and their weakening with lowering temperature (that will cause weakening of the amplitude of the auxiliary wave $Q_1$). In this case the RO will have the non-monotonic temperature dependence of the amplitude, but their frequency and phase will be temperature independent, in accordance with our data (Fig. 3).

## 4. CONCLUSIONS

To conclude, we studied magnetoresistance in the field-induced spin density wave phase ($N=0$) in $(TMTSF)_2PF_6$. We have shown that this supposed to be purely insulating phase, beyond the localized states also contains delocalized ones. The latter occupy two-dimensional pockets, lying in the *a-b* plane, which are quantized in perpendicular magnetic field and give rise to the "Rapid Oscillations". Temperature decreasing do not change the size and orientation of the pockets on the FS, rather, it causes a redistribution of carriers from the delocalized to the localized states. This results in depopulation of the pockets and hence, the disappearance of the RO. Our data agree qualitatively with the theoretical model that considers two coexisting spin density waves with two nesting vectors, respectively. In this model, the second nesting is formed due to the Umklapp processes and the amplitude of the corresponding spin wave should weaken as *T* decreases. Our results clarify the origin of the "Rapid Oscillations" in $(TMTSF)_2PF_6$.

ACKNOWLEDGEMENTS

The work was partially supported by RFBR, EPSRC, RS, INTAS, Programs of the Presidium and Physical Sciences Division of RAS, and the presidential program "The State support of the leading scientific schools".